\documentclass[a4paper,11pt]{article}
\usepackage{amssymb}
\usepackage{jheppub} 
\usepackage{mathrsfs}
\usepackage[T1]{fontenc} 
\usepackage{color}
\newcommand{\vect}[1]{\mbox{\boldmath $#1$}}
\newcommand{\omits}[1]{}
\definecolor{dyellow}{rgb}{1.,0.8,.0}
\definecolor{myblue}{rgb}{.1,.1,.7}
\definecolor{dcyan}{rgb}{.0,.6,.6}
\definecolor{dmagenta}{rgb}{0.6,0.0,0.6}
\definecolor{brown}{rgb}{0.6,0.2,0.}
\definecolor{darkblue}{rgb}{.0,.0,0.5}
\definecolor{darkred}{rgb}{0.75,0.0,0.0}
\definecolor{orange}{rgb}{1.,.6,.0}
\definecolor{dorange}{rgb}{0.8,.4,.0}
\definecolor{green}{rgb}{0.0,1.0,0.0}
\definecolor{lightgrey}{rgb}{0.7,0.7,0.7}
\definecolor{purple}{rgb}{.4,.0,.4}

\begin{document}

\title{The Entropy of Higher Dimensional Nonrotating Isolated Horizons from Loop Quantum Gravity}

\author{Jingbo Wang,}

\author{Chao-Guang Huang}
\affiliation{Institute of High Energy Physics and Theoretical Physics Center for
Science Facilities, \\ Chinese Academy of Sciences, Beijing, 100049, People's Republic of China}

\emailAdd{wangjb@ihep.ac.cn}
\emailAdd{huangcg@ihep.ac.cn}

\abstract{In this paper, we extend the calculation of
the entropy of the nonrotating isolated horizons in 4 dimensional spacetime to that
in a higher dimensional spacetime. We show that the boundary degrees of freedom on
an isolated horizon can  be described effectively by a punctured $SO(1,1)$ BF theory.
Then the entropy of the nonrotating isolated horizon can be
calculated out by counting the microstates.  It satisfies the Bekenstein-Hawking law.  }


\maketitle
\flushbottom


\section{Introduction}

Black hole has been attracting people's attention since a long time ago.
The pioneering works of Bekinstein \cite{bk1}, Hawking \cite{hawk1} and others \cite{bch1}
during the seventies of last century have suggested that black holes have temperature
and entropy. The entropy is given by the famous Bekenstein-Hawking area law
\begin{equation}\label{area-law}
    S=\frac{A}{4G\hbar},
\end{equation}
where $A$ is the area of the event horizon of a black hole.  The entropy
depends on both the Newtonian gravitational and the Planck constants and indicates that
its statistical description
might tell us something profound about quantum gravity. There are many ways
to explain the entropy of a black hole based on different theories,
such as string theory \cite{stringbh1}, loop quantum gravity \cite{asha}.
For a brief review see Ref. \cite{carlip1}.

Unlike the notion of the event horizon of a black hole, which is based on the
global structure of the spacetime \cite{he1}, an isolated horizon is defined
quasilocally as a portion of the event horizon \cite{ash3}.
As expected, the laws of black hole mechanics and thermodynamics can be generalized
to those of isolated horizons \cite{ash2,ash3}.  In particular, the zeroth
and first law of thermodynamics and the area law (\ref{area-law}) for
isolated horizons can also be set up.
The microscopic degrees of freedom on an isolated horizon,
which count for the entropy, are suggested to be described by punctured Chern-Simons
theory \cite{asha} in the framework of loop quantum gravity.

General relativity (GR) in higher dimensional ($D>4$) spacetime has been studied
for almost a century.  The motivations for the study
include Kaluza-Klein theory \cite{Kaluza,Klein}, supergravity theory \cite{NA},
string/M theory \cite{str1,str2,str3}, brane-world scenarios \cite{add1,rs1} and so on.
In higher dimensional gravitational theories, there exist many
 black hole solutions \cite{5d1,5d2}.
Like their 4-dimension partners, black holes in higher dimensional spacetimes also
have temperature and entropy.  Similar to GR, isolated horizons can be introduced in higher
dimensional gravitational theories and the laws of black hole mechanics and
thermodynamics can be generalized, as expected, to those of isolated horizons
\cite{lew1,lew3,lew2}.  How to explain the entropy of the isolated horizons
in higher dimension in the framework of loop quantum gravity \cite{btt1,bn2} is
on the table.  An immediate approach might be to invoke the Chern-Simons theory again.
However, the Chern-Simons theory can only be defined on odd dimensional spacetimes,
which limit its application in higher dimension.

In Ref. \cite{wang1}, we showed that in a 4-dimensional
spacetime, the boundary degrees of freedom on nonrotating isolated horizons can also be
described effectively by another topological field theory---BF theory.
A BF theory can be defined on a spacetime with any dimension, which in this aspect is
the advantage over the Chern-Simons theory.  In the present paper, we extend the results
in Ref. \cite{wang1} to higher dimensional nonrotating isolated horizons.

This paper is organized as follows. In section 2, similar to 4 dimensional case, we derive
the symplectic structure for nonrotating isolated horizons. It can be seen that the
boundary degrees of freedom can be described by a BF theory. In section 3, we quantize
the punctured BF theory and give the corresponding Hilbert space. In section 4, we set
up the boundary condition to relate boundary fields to the bulk fields and calculate the
entropy of the nonrotating isolated horizons. The Bekenstein-Hawking law of the nonrotating
isolated horizons is obtained. Our results are concluded in section 5. In
appendix, we give the detail calculation of the solder field and connection.
Throughout the paper, we use the units of $\hbar=c=1$.


\section{The higher dimensional nonrotating isolated horizons}

The Einstein-Hilbert action can be generalized to $D$ dimensional spacetime
$\mathcal{M}$ \cite{5d2}:
\begin{equation}\label{1a}
    I[g_{\mu\nu}]=\frac{1}{16\pi G}\int_{\mathcal{M}} d^D x\sqrt{-g} R \,.
\end{equation}
It can be written in the first-order form
\cite{lew2}:
\begin{equation}\label{1}
    I[e,A]=-\frac{1}{2\kappa}\int_{\mathcal{M}}F^{IJ}\wedge \Sigma_{IJ},
\end{equation}
where $\kappa=8\pi G$, $e^I$ are covielbein (1-form) fields,
\begin{equation}\label{1b}
    \Sigma_{IJ}=\frac{1}{(D-2)!}\varepsilon_{IJK\cdots N}e^K\wedge \cdots \wedge e^N
\end{equation}
is a $(D-2)$-form, $A^{IJ}$ the $SO(D-1,1)$ connection 1-form, $F_{IJ}$ the curvature 2-form
of $A^{IJ}$, and $I,J$ indices of the Lie algebra of  $\mathfrak{so}(D-1,1)$.
The spacetime region ${\cal M}$ is supposed to be bounded by the initial and final spacelike
hypersurfaces $M_1$ and $M_2$, an isolated horizon $\Delta$ from the inner, and extended to
spatial infinity $i^0$.  All fields are assumed to be smooth and satisfy the standard
asymptotic boundary condition at spatial infinity, $i^0$.

From the first variation of the action (\ref{1}) and variational principle,
we  get the vacuum field equations of gravitation:
\begin{eqnarray}\label{VFE:1}
&&\varepsilon_{IJ\cdots LMN}e^J\wedge \cdots \wedge e^L\wedge F(A)^{MN}=0, \\
&&d_A \Sigma_{IJ}:=d\Sigma_{IJ}-A_I^{\ K}\wedge\Sigma_{KJ}-A_J^{\ K}\wedge\Sigma_{IK}=0,
\label{VFE:2}
\end{eqnarray}
and the symplectic potential density,
\begin{equation}\label{2}
    \theta({\delta})=\frac{1}{2\kappa}\Sigma_{IJ}\wedge {\delta} A^{IJ}.
\end{equation}
The second-order exterior variation will give the symplectic current,
\begin{equation}\label{2a}
    J({\delta}_1,{\delta}_2)=\frac{1}{\kappa}{\delta}_{[2}\Sigma_{IJ}
    \wedge {\delta}_{1]} A^{IJ}.
\end{equation}
The nilpotent of exterior variation, $\vect{\delta}^2=0$, implies $d J=0$.
Applying Stokes' theorem to the integration
$\int_{\mathcal{M}} dJ=0$, we can get the following equation:
\begin{equation}\label{2b}
   \frac{1}{\kappa}\left (\int_{M_2}\delta_{[2}\Sigma_{IJ}\wedge \delta_{1]} A^{IJ}
   -\int_{M_1}\delta_{[2}\Sigma_{IJ}\wedge \delta_{1]} A^{IJ}
   -\int_{\Delta}\delta_{[2}\Sigma_{IJ}\wedge \delta_{1]} A^{IJ}\right )=0 .
\end{equation}
Note that the boundary integral at spatial infinity $i^0$ vanishes by suitable fall-off
conditions \cite{el1}.  We shall see that the
last term in Eq.(\ref{2b}) is a pure boundary contribution, i.e, the symplectic flux
across the isolated horizon $\Delta$ can be expressed as an algebraic sum of
two terms corresponding to the $D-2$ dimensional compact manifold $K_1=\Delta \cap M_1$
and $K_2=\Delta \cap M_2$.

Now let's consider the geometry near the isolated horizon. We adopt the Bondi-like
coordinates $x^{\mu}=(u,r,\zeta^i)$ with coordinate indices $i,j=2,\cdots,D-1$ near the
isolated horizon in Ref. \cite{huang1}.  The isolated horizon $\Delta$ is characterized
by $r=0$.  With  the Bondi-like coordinates, the Bondi-like vielbein vector fields can
be expressed as \cite{huang1,wu2}
\begin{equation} \label{2c}
\left\{ \begin{aligned}
         n^a &= \partial_r \\
         l^a &= \partial_u+U\partial_r+X^i \partial_i\\
         e^a_{\tt A} &=\omega_{\tt A} \partial_r+\xi^i_{\tt A} \partial_i,\qquad \quad
         \mbox{with vielbein indices {\tt A,\,B}}
 =2,\cdots,D-1,
        \end{aligned} \right.
\end{equation}
where $(U,X^i,\omega_{\tt A},\xi^i_{\tt A})$ are functions of $(u,r,\zeta^i)$ and satisfy
$U\triangleq X^i\triangleq \omega_{\tt A} \triangleq 0$.  The symbol $\triangleq$ means
that the equality holds on the isolated horizon $\Delta$. In the following,
we also use $f^{(0)}$ to denote the value of the function $f$ on the isolated horizon.
The covielbein 1-form $n_a,l_a,e^{\tt A}_a$ are
\begin{eqnarray}\label{covielbein}\begin{cases}
  n_a = -du,  \\
  l_a = U du -dr -\xi^{\tt A}_i\omega_{\tt A} (X^idu - d\zeta^i) \qquad \mbox{with $\xi^{\tt A}_i$
   being the inverse of $\xi^i_{\tt A}$},  \\
  e^{\tt A}_a = -\xi_i^{\tt A} (X^i du -d\zeta^i),  \end{cases}
\end{eqnarray}
which satisfy the condition:
$n^a l_a=l^a n_a =-1, e^{\tt A}_a e^a_{\tt B}=\delta^{\tt A}_{\tt B}$,
and others vanish.  The inverse metric reads
\begin{equation}\label{metric}
g^{ab}=-l^a\otimes n^b - n^a \otimes l^b+\delta^{\tt AB}e^a_{\tt A}
\otimes e^b_{\tt B}.
\end{equation}

Define
\begin{equation}\label{3}
    \pi_{\tt A} := e_{\tt A}^a l^b\nabla_b n_a,
\end{equation}
which is related to the angular momentum of the isolated horizon \cite{lew3}.
For the nonrotating isolated horizons, $\pi_{\tt A}\triangleq 0.$
The unknown functions near the nonrotating isolated horizon may be expanded as \cite{wu2}
\begin{equation} \label{4}
\left\{ \begin{aligned}
         U &= \kappa_l r+\frac{1}{2} R^{(0)}_{nlnl} r^2+O(r^3),\\
         \omega_{\tt A} &=\frac{1}{2} R^{(0)}_{n{\tt A}nl} r^2+O(r^3),\\
         X^i &=\frac{1}{2} R^{(0)}_{n{\tt A}nl} \xi^{i(0)}_A r^2+O(r^3),\\
         \xi^i_{\tt A} &=\xi^{i(0)}_{\tt A}-\theta^{(0)}_{\tt AB} \xi^{i(0)}_{\tt B} r+O(r^2),
        \end{aligned} \right.
\end{equation}
where $\kappa_l$ is the surface gravity of the isolated horizon, which is
dependent on the choice of $l$, and $\theta_{\tt AB}:=e_{\tt A}^a
e_{\tt B}^b\nabla_b n_a$.  Due to the zero law of
isolated horizon, $\kappa_l$ is a constant on the horizon.
The asymptotic expansion of the inverse metric near the nonrotating isolated horizon is
then
\begin{equation}\begin{split}\label{5}
    &g^{ur}=1,\quad g^{ui}=0,\\
    &g^{rr}=2U+\delta^{\tt AB} \omega_{\tt A}\omega_{\tt B}=2\kappa_l r+R^{(0)}_{nlnl} r^2+O(r^3),\\
    &g^{ri}=X^i+\delta^{\tt AB} \omega_{\tt A} \xi^i_{\tt B} = \delta^{\tt AB} R^{(0)}_{n{\tt A}nl}
    \xi^{i(0)}_{\tt B} r^2+O(r^3),\\
    &g^{ij}=\delta^{\tt AB}\xi^i_{\tt A} \xi^j_{\tt B} + O(r).
\end{split}\end{equation}

Following the idea of Ref. \cite{wang1} we choose a set of orthogonal vielbein fields
which are compatible with the metric (\ref{metric}) :
\begin{equation}
\label{6}
    e_0^a=-\sqrt{\frac{1}{2}}(\alpha'  n^a  +\frac{1}{\alpha'}  l^a ),\quad
    e_1^a=\sqrt{\frac{1}{2}}(\alpha'  n^a  -\frac{1}{\alpha'} l^a  ),\quad
    e_{\tt A}^a.
\end{equation}
Here $\alpha'(x)$ is an arbitrary function of the coordinates.
$(e_0,e_1)$ with different choices $\alpha(x)$ are related by a Lorentz transformation.
The covielbein fields are given by
\begin{equation} 
    e^0_a=\sqrt{\frac{1}{2}}(\alpha'  n_a +\frac{1}{\alpha'}  l_a ),\quad
    e^1_a=\sqrt{\frac{1}{2}}(\alpha'  n_a -\frac{1}{\alpha'} l_a ), \quad 
    e^{\tt A}_a=e^{\tt A}_{\mu}dx^{\mu}.
\end{equation}

Restricted on the isolated horizon $\Delta$, the 1-form $l$ vanishes, so we have $e^0\triangleq e^1$
(Hereafter, we omit the abstract subscript $a$ for 1-form).
Then the non-zero solder fields on the horizon $\Delta$ satisfy
\begin{equation}\label{15}
    \Sigma_{01}= e^2\wedge e^3 \wedge \cdots \wedge e^{D-1},\quad
    \Sigma_{0\tt A}\triangleq -\Sigma_{1 \tt A}.
\end{equation}
After some straightforward calculation (see Appendix), we can get the following
properties for the $SO(D-1,1)$ connections:
\begin{equation}\label{18a}
    A^{01}\triangleq \kappa_l du+d (\ln\alpha'), \quad A^{0\tt A}\triangleq A^{1\tt A}.
\end{equation}
By Eqs.(\ref{15}) and (\ref{18a}) the integral on the horizon can be reduced to
\begin{equation}\label{19}
\frac{1}{2\kappa}\int_{\Delta} \delta_{[2} \Sigma_{KL}\wedge \delta_{1]} A^{KL}
=\frac{1}{\kappa}\int_{\Delta} \delta_{[2}\Sigma_{01}\wedge\delta_{1]} A^{01},
\end{equation}
since other terms either vanish or cancel with each other.

On the isolated horizon $\Delta$, $\Sigma_{01}$ is just the volume form
of its spatial section.  From the field equation
(\ref{VFE:2}) and the properties (\ref{15}) and (\ref{18a}), it is easy to
show that $d\Sigma_{01}\triangleq 0$, so it is a closed $(D-2)$-form on the horizon $\Delta$.
Locally we can define a $(D-3)$-form $\tilde{B}$ which satisfies
\begin{equation}\label{15a}
  d\tilde{B}=\Sigma_{01} .
\end{equation}

By definition \cite{huang1}, the isolated horizons has the topology of $\mathbb{R}\times K$,
where $K$ is a $(D-2)$-dimensional compact, connected, orientable Riemannian manifold.
Since the topology of the horizon is non-trivial, i.e, the ($D-2$)-th cohomology group
\begin{equation}\label{16}
    H^{D-2}(\mathbb{R}\times K)\cong\mathbb{R},
\end{equation}the $\tilde{B}$ field must satisfy the following condition
\begin{equation}\label{17}
    \oint_{K} |d\tilde{B}|=\oint_{K} |\Sigma_{01}|=a_K,
\end{equation}
where  $a_K$ is the `area' of the horizon, and the second equality comes from the
flux-area relation \cite{blv1}.

From the Eq.(\ref{18a}) we have
\begin{equation}\label{19a}
    dA^{01}\triangleq 0.
\end{equation}
Thus, the integral in Eq.(\ref{19}) can be written as
\begin{equation}\label{BF-SympFlux}
\int_{\Delta} \delta_{[2}\Sigma_{01}\wedge\delta_{1]} A^{01}=
\int_{K_2} \delta_{[2}\tilde B\wedge \delta_{1]}A^{01}-
\int_{K_1} \delta_{[2}\tilde B\wedge \delta_{1]}A^{01}.
\end{equation}
Then, Eq.(\ref{2b}) implies
\[
   \frac{1}{\kappa}\left (\int_{M}\delta_{[2}\Sigma_{IJ}\wedge \delta_{1]} A^{IJ}
   -\int_{K}\delta_{[2}\tilde B\wedge \delta_{1]} A^{01}\right )  \mbox{ is independent of } u,
\]
or
\[
   \frac{1}{\kappa}\int_{M \mbox{ outside } K}\delta_{[2}\Sigma_{IJ}\wedge \delta_{1]} A^{IJ}
    \mbox{ is independent of }u,
\]
which is determined by the bulk only.

Consider an $SO(1,1)$ boost for $(e_0,e_1)$ with $g=\exp(\varsigma)$.
Under the transformation, $A^{'01}=A^{01}-d\varsigma$, and
$\Sigma_{01}$ leaves unchanged. So $A^{01}$ is an $SO(1,1)$ connection,
and $\Sigma_{01}$ is in its adjoint representation.  Eq.(\ref{BF-SympFlux})
is the symplectic flux of an $SO(1,1)$ BF theory across the sections
of the isolated horizon.  Such an $SO(1,1)$ BF theory is what we need
to supplement GR to explain the statistical origin of the area entropy of a
horizon, if the following replacements are made,
 \begin{equation}\label{24}
    B\leftrightarrow \frac{\tilde{B}}{\kappa},\,A\leftrightarrow A^{01},
\end{equation}
as in Ref. \cite{wang1}.


\section{$(D-1)$-dimensional $SO(1,1)$ punctured BF theory}
In $(D-1)$-dimensional spacetime $\Delta$, the action of an ordinary $SO(1,1)$ BF theory
can be written as \cite{bf1}
\begin{equation}\label{22}
    S[B,A]=\int_{\Delta}\textrm{Tr}( B\wedge F(A))=\int_{\Delta}B\wedge dA.
\end{equation}
where $A$ is an $SO(1,1)$ connection field, $F$ its field strength 2-form, and
$B$ a $(D-3)$-form field in the adjoint representation of $SO(1,1)$.
From the action (\ref{22}), we can easily obtain the field equations as
\begin{equation}\label{22a}
    F:=dA=0,\qquad \tilde{F}:=dB=0.
\end{equation}
In the vacuum BF theory, $A$ is a flat connection and $B$-field has a trivial topology.

On the other hand, the field equations for the BF theory we need on the isolated horizon
are
\begin{equation}\label{27}
    F=dA=0,\qquad \tilde{F}=dB=\frac{\Sigma_{01}}{\kappa}.
\end{equation}
Compared with Eq.(\ref{22a}), Eq.(\ref{27}) shows that $B$-field has nontrivial topology
and that the bulk field $\Sigma_{01}$ serves as the source of the $B$ field, locally.
But, $A$ remains a flat connection.

The quantization of the punctured BF theory in $(D-1)$ dimension is similar as in
3 dimension, and we will just summarize the results in \cite{wang1}. Let us assume
that on the spatial slice $K$ there are $n$ punctures denoted by
$\mathcal{P}=\{p_\alpha |\alpha=1,\cdots,n\}$.
For every puncture $p_\alpha$ we associate a $(D-2)$-dimensional bounded neighborhood
$s_\alpha$ which contains it and does not intersect any other.
Denote the boundary of $s_\alpha$ by $\eta_\alpha$. Define the gauge-invariant
functions of the $B$ field
\begin{equation}\label{28}
    f_\alpha=\int_{s_\alpha}dB=\oint_{\eta_\alpha}B .
\end{equation}
The common eigenstates of the corresponding quantum operators $\hat{f}_\alpha$ are
the Dirac distributions $(\{a_p\},\mathcal{P}|\equiv (a_1,a_2,\cdots,a_n|$
characterized by $n$ real numbers $\{a_\alpha, \alpha=1,\cdots,n\}$.
As unbounded self-adjoint operators, the collection $\{\hat{f}_\alpha| \alpha=1,\cdots,n\}$
comprises a complete set of observables in
$\mathcal{H}_K^{\mathcal{P}}\equiv L^2(\mathbb{R}^n)$.
There is a spectral decomposition of $\mathcal{H}_K^{\mathcal{P}}$ with respect to each
$\hat{f}_\alpha$, i.e,
\begin{equation}\label{29}
    (\{a_p\},\mathcal{P}|\hat{f}_\alpha=(\{a_p\},\mathcal{P}|a_\alpha.
\end{equation}


\section{The entropy of isolated horizon}

In order to calculate the entropy of an isolated horizon, we
consider the system described by the action (\ref{1}) for the bulk ${\cal M}$ plus
an $SO(1,1)$ BF theory for the isolated horizon
as the internal boundary of ${\cal M}$, whose $B$-field has nontrivial topology.
In loop quantum gravity approach, only the horizon degrees of freedom contribute
to the black hole entropy.  Hence, the bulk degrees of freedom need to be traced out.
We can construct a density matrix $\rho_{BH}^{}$ for such a system and assume the
system is in a maximally mixed state.  The entropy can
be given by the von Neumann formula
\begin{equation}\label{29a}
    S_{BH}=-\textrm{Tr}(\rho_{BH}^{}\ln \rho_{BH}^{}),
\end{equation}or equivalently,
\begin{equation}\label{29b}
    S_{BH}=\ln \mathcal{N}_{BH},
\end{equation}where $\mathcal{N}_{BH}$ is the total number of the states in the
horizon Hilbert space that satisfy some constraints.  The goal of this section is
to compute this number to give the entropy.

Following the paper \cite{wang1}, the classical boundary condition relating the boundary and bulk
fields are chosen as
\begin{equation}\label{30}
    \tilde{F}(x)\circeq \Sigma_{01}(x)/\kappa,
\end{equation}
or its integral form:
\begin{equation}\label{31}
    \oint_K \tilde{F}(x)\circeq\frac{1}{\kappa} \oint_K \Sigma_{01}(x),
\end{equation}
where $\tilde{F}$ is the exterior differential of the $B$-field on the boundary,
$\Sigma_{01}$ is the canonical momentum density conjugate to $A^{01}$ in the bulk, and
$\circeq$ means that the equality is valid when the limit of $\Sigma_{01}$ at
the spatial slice $K$ of the isolated horizon is taken.
The right hand side of the equality is proportional to the flux
of $\Sigma_{01}$ across $K$, $Fl_{01}(K):=\oint_K \Sigma_{01}(x)$,
which is well defined in loop quantum gravity \cite{btt4}.  The left hand side is
the volume integral of $\tilde F$ over $K$, which is just
the sum of the observables (\ref{28})
in the loop-quantized version of the BF theory.

The Hilbert space for the bulk theory in higher dimension can be constructed in the
approach proposed by Bodendorfer and his collaborators \cite{btt2,btt3,btt4,btt5}.
The key point is that one can proceed in the phase space for the Euclideanization
of the $D$-dimensional spacetime, which consist of an $SO(D)$ connection $A_{a}^{IJ}$
and its conjugate momentum $\pi^{a}_{IJ}$.
Besides the usual Gauss constraint, the spatial diffeomorphism constraint and Hamiltonian
constraint, there is a simplicity constraint. In the quantum theory, the
simplicity constraint can be implemented on the links of a spin-network by restricting
the representations of the $SO(D)$ to be of class 1, so that their highest weight vector
$\vec{\lambda}$ is determined by a single non-negative integer $\lambda$ as
$\vec{\lambda}=(\lambda,0,\cdots,0)$ \cite{fkkp1}. Under the
representation, the `area' $Ar[S]$ for a
$(D-2)-$hypersurface $S$ can be constructed as $\sqrt{\mbox{`flux squared'}}$,
i.e. \cite{btt4},
\begin{equation}\label{31b}
    Ar[S]:=\sum_U \sqrt{\frac{1}{2} Fl_{IJ}(S_U) Fl^{IJ}(S_U)},
\end{equation}
where $S=\bigcup S_U$ is a partition of the hypersurface $S$ by a set of
closed sets $\{S_U\}$ with each $S_U$ containing, at most, one puncture,
and $Fl_{IJ}(S_U)$ is the flux through $S_U$ which can be quantized properly.

Such a construction has been used in the study of the isolated horizons in higher dimensional
spacetime \cite{btt1}. For $2n$-dimensional spacetime, the degrees of freedom on the
$(2n-1)$-dimensional isolated horizon $\Delta$ can be described by an $SO(2n)$
Chern-Simons theory.  Unfortunately, the non-Abelian Chern-Simons
theory with $n>1$ has local degrees of freedom, which would result in the
divergence of the entropy. In order to avoid the problem,
a stronger boundary condition is proposed.  The stronger boundary
condition relates
the Chern-Simons connection on the boundary to a hybrid
connection on the bulk.
In contrast, there is no such a problem with BF theory, since a BF theory
has no local degree of freedom even in higher dimension \cite{bf1}.

The eigenvalues of the flux operator of the canonical
momentum conjugate to $A^{01}$ across an arbitrary  $(D-2)$-dimensional
closed, connect, oriented, spacelike hypersurface $S$ in $M$ has the form
\begin{equation}\label{32}
    \oint_S  \hat{\Sigma}_{01}(x)  |{m_\alpha,\cdots}>
=8\pi G \beta\sum_{\alpha} m_\alpha |{m_\alpha,\cdots}>,
\end{equation}
where $\alpha$ indicates the $\alpha$-th puncture of a spin-network eigenstate on
$S$, which coincides with punctures on the boundary BF theory,
$m_\alpha \in \{-\lambda_\alpha,-\lambda_\alpha+1,\cdots,\lambda_\alpha\}
\subset \mathbb{Z}$
is the quantum number associated with the flux operator, $|{m_\alpha,\cdots}>$ represent
an $SO(D)$ spin network state in the bulk, and $\cdots$
represents other quantum numbers that character the state,
such as the quantum number associated with the intertwinner operator.

The quantum version of the boundary condition (\ref{31}) reads
\begin{equation}\label{33}
(\textrm{Id} \otimes \oint_{s_\alpha} \hat{\tilde{F}}
-\frac{1}{8\pi G}\oint _{s_\alpha} \hat{\Sigma}_{01}\otimes \textrm{Id} )(\Psi_v \otimes \Psi_b)=0,
\end{equation}where $\textrm{Id}$ means the identity operator, $\Psi_v$ and $\Psi_b$
bulk and boundary states, respectively,
$s_\alpha$ the $(D-2)$-dimensional bounded neighborhood associated to
the puncture $p_\alpha$ as before.

From Eq. (\ref{33}), we can get the relation between
the eigenvalues of $\hat{\tilde{F}}$ and $\hat\Sigma_{01}$:
\begin{equation}\label{34}
    a_\alpha=\beta m_\alpha,\quad m_\alpha \in  \mathbb{Z}.
\end{equation}
In other words, $a_\alpha$ is no longer any real number, but takes discrete values.

The eigenvalues of the flux-area operator which appears in the quantum version
of Eq. (\ref{17}) has the form
\begin{equation}\label{flux-area-eigenvalue}
    \oint_S \parallel \hat{\Sigma}_{01}(x) \parallel |{m_\alpha,\cdots}>
=8\pi G \beta\sum_{\alpha} |m_\alpha| |{m_\alpha,\cdots}>.
\end{equation}
Then, a global constraint appears from the equation (\ref{17}) and (\ref{24}):
\begin{equation}\label{35}
    \sum_\alpha |a_\alpha|=a_K/(8\pi G).
\end{equation}
Similar to the `flux-area' constraint in Ref.\cite{blv1},
(\ref{35}) is called flux constraint.
With the Eq.(\ref{34}), the constraint can be reformulated as
\begin{equation}\label{35a}
  \sum_\alpha |m_\alpha|=a_K/(8\pi\beta G),\quad m_\alpha \in \mathbb{Z}.
\end{equation}

So the number of compatible states is given by
\begin{equation}\label{36}
     \mathcal{N}_{BH}=\sum_{n=1}^{n=a} C_{a-1}^{n-1} 2^n=2\times 3^{a-1}
\end{equation}
where $a=a_K/(8\pi \beta G) \in \mathbb{N}$.
The entropy is given by
\begin{equation}\label{38}
    S_{BH}=\ln\mathcal{N}_{BH}=a\ln3+\ln \frac{2}{3}=\frac{ \ln3}{2\pi \beta}\frac{a_K}{4G}+\ln \frac{2}{3}.
\end{equation}
Finally we get the area law.
Besides, if we set $\beta=\ln 3/(2\pi)$, we also get the famous coefficient $1/4$. Note that this value is dimension-independent.
Compared with the case in 4 dimension \cite{wang1}, this parameter has an additional
factor 2. This is due to the fact that in higher dimension we use the group $SO(D)$
which only has integer representation, and in 4 dimension we use $SU(2)$ which
can have half-integer representation.  If the present procedure is applied to $D=4$ case,
in which $SO(4)$ group instead of $SU(2)$ group is used, the $\beta$ parameter will
be different, but the spectrum of the area operator remains the same.
Whether there exist any physical process which
can distinguish the two approaches is an interesting problem.

As the same as in 4 dimensional spacetime, the entropy we get has a constant correction
term besides the leading area term.   The `zero-point
entropy'  first appears as the quantum correction for the area law
of entropy in Ref. \cite{huang2,huang3}.


\section{Discussion}

In this paper, we calculate the entropy of nonrotating isolated horizon in higher
dimensional spacetime following the standard procedure in loop quantum gravity \cite{ash1}.
From the first-order action (\ref{1}), the presymplectic current is obtained.  The current through
the isolated horizon $\Delta$ can be reformulated as the difference
across its final and initial spatial sections.
Then the symplectic form on a $(D-2)$-dimensional hypersurface is acquired.
The degrees of freedom on the cross section of the isolated horizon can be
described by a punctured $SO(1,1)$ BF theory with the symplectic form.
The result is the same as in 4 dimensional spacetime.

Notice that in the present calculation of the entropy of isolated horizon,
the area constraint is not used, which plays an important role in the Chern-Simons
theory approach.  Instead, the flux constraint (\ref{35}) is used.
Classically, the area of the horizon equals
its flux-area, which can
be seen from the equation (\ref{17}). But at quantum level, they
correspond to two different operators: (\ref{31b}) and (\ref{flux-area-eigenvalue}), respectively.
The eigenvalues of the `area' operator are given by \cite{btt4}
\[
8\pi G \beta \sum_{\alpha}\sqrt{\lambda_{\alpha}(\lambda_{\alpha}+D-2)},
\quad \lambda_{\alpha}\in \mathbb{N},
\]
where $\beta$ is a parameter analogous to the Barbero-Immirzi parameter in 4 dimension.
Those eigenvalues are obviously different from that of the  flux-area operator
(\ref{flux-area-eigenvalue}).
We use the flux constraint because in a loop quantization of a generalized gravity the flux operator
turns out to measure the Wald entropy \cite{bn1}.

The starting point of our approach is the first-order action (\ref{1}) which has
$SO(D-1,1)$ as the gauge group.  On the isolated horizon, we make a gauge fixing
into its subgroup $SO(1,1)$.  Taking this fact into account, a punctured BF with gauge
group $SO(1,1)$ is chosen to describe the boundary degrees of freedom.
But the quantum states in the bulk are well presented only from the Hamiltonian
framework with gauge group $SO(D)$ \cite{btt2,btt3}, since the non-compact $SO(D-1,1)$
group is not suitable for loop quantized.
The mismatch of gauge group $SO(D-1,1)$  and
$SO(D)$ in the bulk, in particular, near the isolated horizon
maybe cause some confusion.  In fact, in the calculation of the entropy,
the bulk degrees of freedom are traced out, and what we count is just the number of
states compatible to the boundary states.   The bulk theory enter into the calculation
through the form of the eiganvalues for flux operator (\ref{32}).  Therefore,
the calculation is still reasonable.

In Ref. \cite{btt1}, due to the difficulty of the non-Abelian Chern-Simons theory in
higher dimension, it is suggested to
use $n^{[I}\tilde{s}^{J]}$ as horizon degrees of freedom for all higher dimensions.
It is worth to compare the entropy of a horizon obtained in the two different
approaches.


\appendix
\section{Solder field and connection on the isolated horizon}
The connections $A_I^{\ J}$ which are adapted with the vielbein $e_I$ are given by
\begin{equation}\label{40}
    (A_I^{\ J})_a=(e_I)^c [\partial_a (e^J)_c-\Gamma^b_{ac}(e^J)_b].
\end{equation}
Since what we concern is the connection restricted on the isolated horizon with $r=0$,
which has no $dr$ component, we only need to calculate it in the infinitesimal neighborhood
of the horizon, so that the $e^I$ and the Christoffol symbol $\Gamma^a_{bc}$ are kept to
the zero order of $r$ and metric to the first order of $r$.

First we choose the parameter $\alpha'(x)=1$, and other cases can change into this case
through a Lorentz transformation.
Thus, from Eqs. (\ref{6}), (\ref{2c}) and (\ref{covielbein}), one has
\begin{equation}
\label{41}
    e_0=-\sqrt{\frac{1}{2}}(\partial_u +\partial_r)+O(r),
\quad e_1=\sqrt{\frac{1}{2}}(-\partial_u +\partial_r)+O(r),\quad
    e_{\tt A}=\xi^i_{\tt A} \partial_i+O(r),
\end{equation}
and
\begin{equation}
\label{42}
    e^0=-\sqrt{\frac{1}{2}}(du+dr)+O(r),\quad e^1=\sqrt{\frac{1}{2}}(-du +dr)+O(r),\quad
    e^{\tt A}=\xi_i^{\tt A} dx^i+O(r).
\end{equation}
On the horizon $r=0$,
\begin{equation}\label{43}
    e^0\triangleq e^1\triangleq -\sqrt{\frac 1 2}du
\end{equation}
and
\begin{equation}\label{44}
    \Sigma_{01}= e^2\wedge e^3 \wedge \cdots \wedge e^{D-1},\quad
    \Sigma_{0\tt A}\triangleq -\Sigma_{1\tt A}.
\end{equation}

Next we calculate the connection $(A_0^{\ 1})_a =(e_0)^c [\partial_a (e^1)_c-\Gamma^b_{ac}(e^1)_b]$
[or $(A_0^{\ 1})_\mu dx^\mu
=e_0^\sigma (\partial_\mu e^1_\sigma-\Gamma^\nu_{\mu\sigma}e^1_\nu)dx^\mu$]
restricted on the horizon $r=0$.
From the expression (\ref{41}) and (\ref{42}) we can see that
$\nu, \sigma$ can only take values corresponding to $u,r$.
For $ \mu=u$, the nonzero Christoffol symbols are
$\Gamma^u_{uu}\triangleq \Gamma^r_{ur}\triangleq\kappa_l$, so we can get
 $(A_0^{\ 1})_u\triangleq-\kappa_l$.
For $ \mu=i$, with same method, it can be
shown that $(A_0^{\ 1})_i\triangleq0$.
So in this case, we can get
 \begin{equation}\label{45}
    A^{01}\triangleq\kappa_l du.
 \end{equation}
 For $\alpha'(x)\neq 1$, which relate to our case with a Lorentz transformation,
 the connection transforms into $ A^{01}\triangleq \kappa_l du+d (\ln\alpha')$,
 which is the first result of (\ref{18a}).

Finally, the second result of (\ref{18a}) is equivalent to
$(A_0^{\ \tt A})_a\triangleq -(A_1^{\ \tt A})_a$, or
\begin{equation}\label{46}
  [(A_0^{\ \tt A})_\mu+(A_1^{\ \tt A})_\mu]dx^\mu=(e_0^\sigma+e_1^\sigma)
  (\partial_\mu e^{\tt A}_\sigma-\Gamma^\nu_{\mu \sigma}e^{\tt A}_\nu)dx^\mu\triangleq 0.
\end{equation}
Again, $\sigma$ can only take values corresponding to $u, r$.
For $\sigma=r$, we already have $e_0^r+e_1^r\triangleq 0$.
For $\sigma=u$, we have
 \begin{equation}\label{47}
  (e_0^u+e_1^u) (\partial_\mu e^{\tt A}_u-\Gamma^\nu_{\mu u}e^{\tt A}_\nu)
  \triangleq \sqrt{2}\Gamma^i_{\mu u}e^{\tt A}_i\triangleq 0,
  \end{equation}since
\begin{equation}\label{48}
   \Gamma^i_{\mu u}=\frac{1}{2} g^{i\nu}(g_{\mu\nu,u}+g_{u\nu,\mu}-g_{\mu u,\nu})
     \triangleq\frac{1}{2} g^{ij}g_{kj,u}\delta^k_{\mu}
 \triangleq 0.
\end{equation}
In the last step, $d\Sigma_{01}\triangleq 0$ has been used.
So, we complete the proof of the results (\ref{18a}).

\acknowledgments
One of the authors (Wang) would like to thank Prof. Yongge Ma for helpful discussion.
This work is supported by National Natural Science Foundation of China under the grant
11275207.

\bibliography{btz3-140903h1}
\end{document}